\documentclass[%
 reprint,
 amsmath,amssymb,
 aps,
]{revtex4-2}
\usepackage[utf8]{inputenc}
\usepackage{xcolor}
\usepackage{comment}
\usepackage{graphicx}% Include figure files
\usepackage{dcolumn}% Align table columns on decimal point
\usepackage{bm}% bold math
\begin{document}

\preprint{APS/123-QED}

\title{Tuning magnetic exchange interactions in 2D magnets: the case of 
CrGeX$_3$ (X = Se, Te) and Janus Cr$_2$Ge$_2$(Se,Te)$_3$ monolayers}% Force line breaks with \\

\author{Gabriel Martínez-Carracedo$^{1,2}$}
\author{Amador García-Fuente$^{1,2}$}
\author{László Oroszlány$^{3,4}$}
\author{László Szunyogh$^{5,6}$}
\author{Jaime Ferrer$^{1,2}$}

\affiliation{
$^1$Departamento de Física,  Universidad de Oviedo,  33007 Oviedo, Spain\\
$^2$Centro de Investigación en Nanomateriales y Nanotecnología, Universidad de Oviedo-CSIC, 33940, El Entrego, Spain\\
$^3$Department of Physics of Complex Systems, Eötvös  Loránd University, 1117 Budapest, Hungary\\
$^4$Wigner Research Centre for Physics, H-1525, Budapest, Hungary\\
$^5$Department of Theoretical Physics, Institute of Physics, Budapest University of Technology and Economics, 
M\H{u}egyetem rkp. 3., H-1111 Budapest, Hungary \\
$^6$HUN-REN-BME Condensed Matter Research Group, Budapest University of Technology and Economics, 
M\H{u}egyetem rkp. 3., H-1111 Budapest, Hungary
}

\begin{abstract}
We present a computational study to explore the potential of different experimental approaches to tune the magnetic interactions in two-dimensional van der Waals magnets. 
We selected CrGeSe$_3$, CrGeTe$_3$, and Janus Cr$_2$Ge$_2$(Se,Te)$_3$ monolayers as case studies  and calculated the full exchange tensors among all relevant atomic pairs and analyze their dependence on different external parameters, such as biaxial and uniaxial strain, as well as gate voltage. We focus particularly on interactions that emerge or vanish due to changes in the system's symmetry, especially under uniaxial strain and gate voltage. We find that biaxial and uniaxial strains significantly modify isotropic exchange couplings, which can lead to a transition from a ferromagnetic to an antiferromagnetic phase, while a gate voltage induces  Dzyaloshinskii–Moriya interactions, forming a vortex pattern whose chirality is determined by the sign of the electric field. 
The electric dipole moment of the Janus material
is large, raising the possibility of multiferroic behavior. The
polarizability is similar for the three compounds.

\end{abstract}


\maketitle

%\tableofcontents

\section{Introduction}
Since the initial experimental report of magnetism in two-dimensional (2D) materials in 2017 \cite{Huang2017,Gong2017}, the interest of the scientific community in van der Waals (vdW) magnets has greatly increased. On the one hand, they promise to be strong candidates for future magnetic memory technologies like MRAM-based devices \cite{Yang2022,RevModPhys.96.015005} and due to their layered exfoliation behavior they can successfully facilitate the miniaturization process. On the other hand, the magnetic domain structure found in 2D vdW magnets \cite{Sun2021,Yang2022FGT} might open the way toward developing domain wall memory technology in 2D materials \cite{Kumar2022}. \\
Long-range magnetic order appears in 2D vdW magnets because the magnetic anisotropies arising from spin-orbit coupling (SOC) are large enough to overcome thermal and quantum fluctuations \cite{PhysRevLett.17.1133}. However, the experimental Curie temperatures ($T_c$) of these materials remain relatively low: 35 K for CrI$_3$ monolayer \cite{Huang2017} or 175 K for 10 layer thick Fe$_3$GeTe$_2$ \cite{Roemer2020}.
More promisingly, SOC can lead to non-collinear magnetic configurations,
such as helical states \cite{Bode2007,PhysRevB.100.214406} or spin textures 
\cite{PhysRevB.106.054426,TRAN2024115799} that have been suggested to be candidates for  data storage 
    \cite{WANG2022169905}. In most cases these are caused by the antisymmetric exchange, i.e., the Dzyaloshinskii-Moriya (DM) interaction. In centrosymmetric materials like NiI$_2$ monolayers, where the DM interaction vanishes because of inversion symmetry, skyrmionic structures can be stabilized with the aid of symmetric exchange \cite{Amoroso2020}. Other magnets that have been proposed to host non-collinear magnetic textures are Janus materials. The interest in Janus magnets lies in the fact that due to the lack of inversion symmetry it is easier to induce non-collinear magnetic textures triggered by DM interactions \cite{CGST}, providing a solid platform for skyrmion-based devices in the near future.  Examples of Janus magnets that have been studied previously include Cr$_2$(I,X)$_3$ (X = Br, Cl, F) \cite{Sun2022,PhysRevB.101.060404}, Ni(X,Y) (X,Y = I, Br, Cl) \cite{PhysRevMaterials.7.054006}, MnBi$_2$X$_2$Te$_2$ (X = Se, S) \cite{PhysRevB.103.104403} or Cr(Y,X) (Y = S, Se, Te; X = Cl, Br, I) \cite{Hou2022}. To date, the synthesis of Janus monolayers has been limited, with only a few examples, such as SPtSe \cite{Sant2020} and MoSeS \cite{Jang2022}.

The recently developed computational tool {\sc grogu} \cite{PhysRevB.108.214418,Grogu} enables us to quantify the bilinear tensorial exchange interactions between any two atomic magnetic
entities from first principles calculations, well beyond the capability
of methods based on energy differences \cite{PhysRevB.84.224429}. In this work, we take advantage of this tool to analyze how different parameters, which can be controlled experimentally, affect these interactions and could even lead to a change in the magnetic configuration of a 2D vdW magnet. In particular, we consider two types of effects. One is the change in the bond distances and angles 
between magnetic atoms, which can be controlled by the application of strain. As we show here, strain can have a strong effect on the magnitude and even the sign of the magnetic interactions, eventually inducing a phase transition between ferromagnetic (FM) and antiferromagnetic (AFM) states. A second approach consists of changing the system's symmetry, which can be achieved by atomic changes in the structure, uniaxial strain, or the application of a gate voltage. Certain magnetic interactions are forbidden in high-symmetry structures, while lowering the
symmetry of the system can allow some of those interactions to appear. 
We show that large DM interactions can be activated and tuned by uniaxial strain, and especially by a gate voltage. Such effects then can promote the appearance of non-collinear magnetic structures with a simple experimental knob. In fact, experiments proved that a gate voltage can tune the chirality of skyrmions in centrosymmetric monolayers such as CrGeTe$_3$ (CGT) \cite{Han2024}.

\begin{figure*}[ht!]
    \centering
    \includegraphics[width=2\columnwidth]{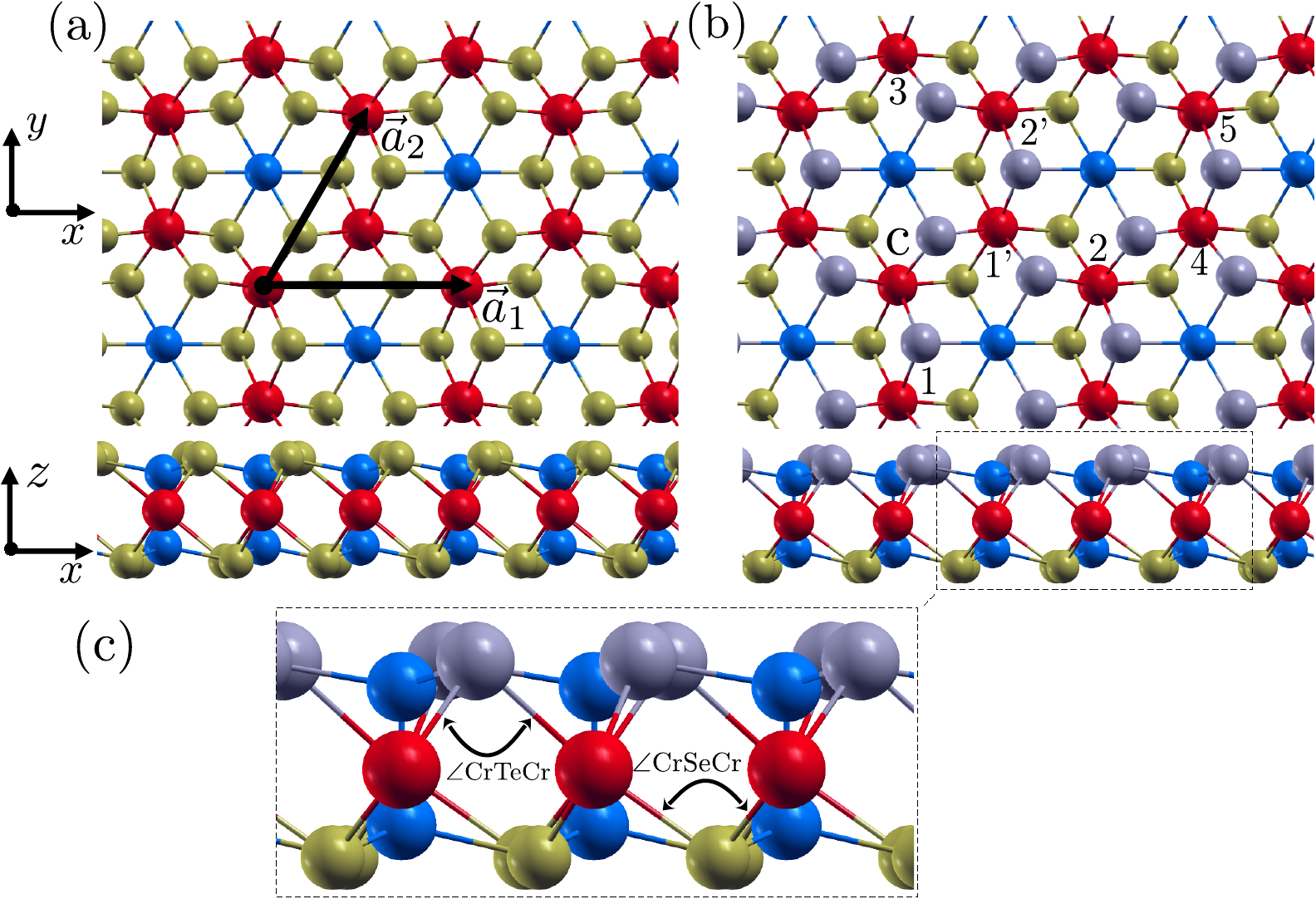}
    \caption{Top and side views of (a) the CGS and (b) the Janus CGST structures. Cr, Ge, Se and Te atoms are depicted in red, blue, yellow and grey, respectively. Lattice vectors $\vec{a}_1$ and $\vec{a}_2$ are shown in (a). Labeling of the nearest Cr neighbors with respect to a reference atom $c$ is also shown in (b). Cr-X-Cr (X = Se or Te) bond angles are shown in (c).}
    \label{Structure}
\end{figure*}

To illustrate how magnetic transitions can be triggered, we study the monolayers of CrGeSe$_3$ (CGS) and CGT. The magnetic, electronic, and magnonic properties of these vdW materials have been widely studied \cite{Ren_2020,Hao2021,Chen2022,doi:10.1126/sciadv.abi7532,PhysRevB.98.125127,Liu2021}, and interest in them has increased in part because of the new 2D physics they can offer.  CGS and CGT monolayers possess a rhombohedral structure and are characterized by the space group $R\bar{3}$. Their minimal unit cell contains 10 atoms and the structure is depicted in Fig.~\ref{Structure}(a). We also study a Janus version of this structure, where the top external layer is formed by Te and the bottom external by Se atoms as depicted in Fig.~\ref{Structure}(b).
This system corresponds to the Cr$_2$Ge$_2$(Se,Te)$_3$ (CGST) stoichiometry. Phonon spectra calculations have shown that these lattice structures are stable \cite{CGST,PhysRevApplied.12.014020}. Recently, Monte-Carlo calculations have shown that Janus CGST can stabilize skyrmionic states \cite{CGST}. \\
This article aims to provide a comprehensive understanding of how symmetry breaking through external perturbations affects the exchange parameters, serving as a clear guide for future control and tuning of the magnetic properties of 2D magnets. In section \ref{Sec_model} we define the parts of the spin Hamiltonian used to model the magnetic interactions and determine which ones vanish due to the symmetry of the systems under consideration. In section \ref{Sec_method} we give the details of the calculations. In section \ref{Sec_results} we present and explain the results obtained when biaxial and uniaxial strain, as well as a gate voltage are applied. 
Finally, in Section \ref{Sec_conclusions} we summarize and draw our conclusions.

\section{Spin model and symmetry constraints}\label{Sec_model}

Most magnetic materials with localized magnetic moments can be described by the classical Heisenberg Hamiltonian:
  \begin{equation}
 H(\{ {\bf e}_i\})
=\,\frac{1}{2}\,\sum_{i\neq j}\,{\bf e}_i\,{\mathcal J}_{ij}\,{\bf e}_j+\sum_i\,{\bf e}_i\,{\mathcal K}_i\,{\bf e}_i \, ,
     \label{HeisenbergEq}
 \end{equation}
\noindent
where ${\bf e}_i$ is a unit vector pointing along the spin-only magnetic momentum of the magnetic atom %entity 
located on site $i$, ${\mathcal K}_i$ is the on-site anisotropy matrix on site $i$, and ${\mathcal J}_{ij}$ is the exchange 
tensor that can be written as the sum of a symmetric and an antisymmetric part:
\begin{equation}
\begin{split}
\label{Js+Ja}
&{\mathcal J}_{ij}={\mathcal J}_{ij}^s+{\mathcal J}_{ij}^a= \\[5pt] =&\begin{pmatrix} J^{xx}_{ij} & S^z_{ij} &  S^y_{ij} \\  S^z_{ij} & J^{yy}_{ij} & S^x_{ij} \\ S^y_{ij} &  S^x_{ij} & J^{zz}_{ij} \end{pmatrix}+
\begin{pmatrix}   0    & D^z_{ij} & -D^y_{ij} \\ -D^z_{ij} &   0    & D^x_{ij} \\ D^y_{ij} & -D^x_{ij} &    0   \end{pmatrix} \, .
\end{split}
\end{equation}
The DM vectors are then defined as ${\bf D}_{ij}= (D^x_{ij},D^y_{ij},D^z_{ij})$. It is common to write the spin-Hamiltonian \eqref{HeisenbergEq} in the following form,
%extract the isotropic part from ${\mathcal J}_{ij}^s$ leading to:
\begin{equation}
\begin{split}
    H(\{ {\bf e}_i\})&=\,\frac{1}{2}\,\sum_{i\neq j}\,J_{ij}^H\,{\bf e}_i\cdot{\bf e}_j+
\frac{1}{2}\,\sum_{i\neq j}\,{\bf e}_i\,{\mathcal J}_{ij}^S\,{\bf e}_j+\\+&\,\frac{1}{2}\,\sum_{i\neq j}\,{\bf D}_{ij}\cdot({\bf e}_i\times{\bf e}_j)+
\sum_i\,{\bf e}_i\,{\mathcal K}_i\,{\bf e}_i \, ,
\end{split}
    \label{HeisenbergEq2}
\end{equation}

\noindent
where $J_{ij}^H=\frac{1}{3}\text{Tr} J_{ij}^s$ is the isotropic Heisenberg coupling and ${\mathcal J}_{ij}^S={\mathcal J}_{ij}^s- J_{ij}^H
%\frac{1}{3}\text{Tr} {\mathcal J}_{ij}^s 
\, {\mathcal I}$ is a traceless symmetric matrix.

The knowledge of the exchange constants is crucial for computing and predicting %physical quantities such as 
the order-disorder transition temperature $T_c$, the magnon spectra or the spin pair correlation functions. Moreover, external control over these constants through electric fields or strain can result in magnetic phase transitions 
\cite{PhysRevB.107.035432,Xu2020,PhysRevB.91.140405,Cenker2022}, generate skyrmionic textures
\cite{Sun2023,PhysRevApplied.19.024064} or tune $T_c$ \cite{siskins2022}.

\begin{figure}[h]
    \centering
    \includegraphics[width=\columnwidth]{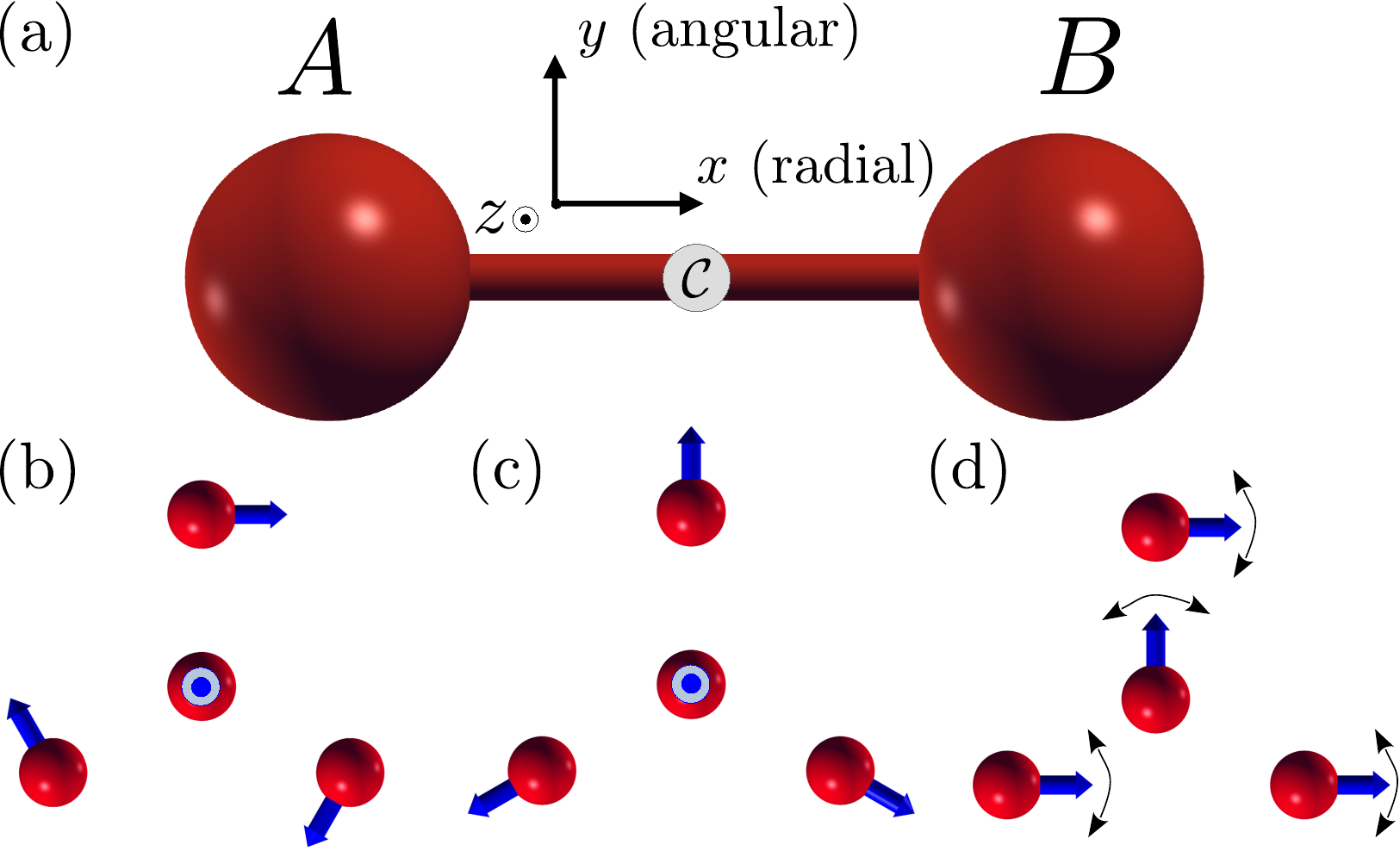}
    \caption{(a) Schematic picture of the local frame of reference, where the bond connecting the $A$ and $B$ atoms is parallel to the $x$-axis. The positive $z$ direction points towards the reader. Symmetries and selection rules are computed with respect to the center $\mathcal{C}$. (b-d)  
    Sketch of non-collinear textures induced by DM vectors at first nearest neighbors relative to the central atom. Magnetic moments are represented by blue arrows. (b) A vortex pattern is formed around the central atom if only a radial DM component is present.  (c) A radial vortex pattern is formed around the central atom if only an angular DM component is present. In panels (b) and (c), the spin in the central atom is pointing towards the reader. (d) An in-plane relative 90$^{\circ}$-degree orientation among first neighbors is boosted when the nearest neighbor DM vectors are normal to the $xy$ plane. The thin curved arrows demonstrate that this magnetic pattern can be rotated in the plane at zero energy cost.}
    
    \label{bondDM}
\end{figure}

In our calculations, ${\mathcal J}_{ij}$ is computed at the local frame of reference shown in Fig. \ref{bondDM}(a), while the corresponding atomic pairs $ij$ are labeled as indicated in Fig. \ref{Structure}(b). In this figure, each pair $ij \in \{c1, c2, c3, c4 \}$ exhibits a different point group symmetry, that leads to different selection rules for the exchange constants. In the Appendix, we summarize the selection 
rules for the exchange constants when inversion symmetry ($\Pi$), mirror $\alpha\beta$-plane symmetry  ($\sigma_{\alpha\beta}$), and $C_2$ symmetry around the $\alpha$-axis ($C_{2\alpha}$) (where $\alpha$, $\beta$ = $x$, $y$, and $z$) are present with respect to the center of a given bond.  We apply
these rules to identify the non-zero exchange parameters from the corresponding symmetries for the
CGS and CGST structures, see Table \ref{tablesymm}. The Janus structure of CGST breaks $\Pi$, $\sigma_{xy}$, $C_{2x}$ and $C_{2y}$ symmetries that are present in CGS and CGT. For CGST, the first and third neighbors have non-zero $D^y$ and $D^z$ values, which vanish for CGS and CGT. We will demonstrate that it is also possible to induce $D^y$ in CGS and CGT between first neighbors by a gate voltage that breaks inversion symmetry. Moreover, uniaxial strain breaks $C_{2y}$ symmetry with respect to the $c2$' bond, inducing a $D_{c2\text{'}}^y$ component. In Fig. \ref{bondDM}(b-d) we show a sketch of how different DM vector components induce non-collinear magnetic configurations. 

\begin{table*}[ht!]
\small
  \caption{Symmetries and non-zero exchange parameters for the first five nearest neighbors in CGS / CGST structures following the local frame of reference  depicted in Fig \ref{bondDM}(a).  For simplicity, we removed $J^{xx}$, $J^{yy}$ and $J^{zz}$ from the list of non-zero exchange parameters as there are no selection rules for these matrix elements. }

  \begin{tabular*}{\textwidth}{@{\extracolsep{\fill}}cccc}
    \hline
    Bond &  Symmetries centered on $\mathcal{C}$&  Non-zero exchange parameters\\
    \hline
     c1      &$\Pi$, $\sigma_{yz}$, $C_{2x}$ / $\sigma_{yz}$&$S^{x}$ / $D^{y}$, $D^{z}$, $S^{x}$\\
     c2      & $C_{2y}$ / $\emptyset$& $D^{x}$, $D^{z}$, $S^{y}$ / $D^{x}$, $D^{y}$, $D^{z}$, $S^{x}$, $S^{y}$, $S^{z}$ \\
     c3      &$\Pi$, $\sigma_{yz}$, $C_{2x}$ / $\sigma_{yz}$&$S^{x}$ / $D^{y}$, $D^{z}$, $S^{x}$\\
     c4      & $\Pi$ / $\emptyset$& $S^{x}$, $S^{y}$, $S^{z}$ / $D^{x}$, $D^{y}$, $D^{z}$, $S^{x}$, $S^{y}$, $S^{z}$\\
     c5& $C_{2x}$ / $\emptyset$& $D^{x}$, $S^{x}$ / $D^{x}$, $D^{y}$, $D^{z}$, $S^{x}$, $S^{y}$, $S^{z}$\\
    \hline
  \end{tabular*}
  \label{tablesymm}
\end{table*}

\section{Computational method}\label{Sec_method}

We %are able to 
compute all terms in the spin-Hamiltonian given in Eq. (\ref{HeisenbergEq2}) by a two-step process. First, we obtain the electronic Hamiltonian based on Density Functional Theory 
(DFT) \cite{HK,KS} by using the {\sc siesta} code \cite{siestapaper}. This code has the
advantage that electronic wave functions are expanded in a basis of pseudo-atomic orbitals,
from which it is easy to identify the atomic magnetic moments. In as second step, the DFT Hamiltonian is used as input of the package {\sc grogu} \cite{PhysRevB.108.214418}, which extracts the exchange and anisotropy constants of the system based on the magnetic force theorem \cite{liechtenstein87,Udvardi2003}.

Our DFT calculations employed the Generalized Gradient Approximation (GGA) in the 
scheme of Perdew-Burke-Ernzerhof (PBE) \cite{PBE} for the exchange-correlation potential. A 
10$\times$10$\times$1 Monkhorst-Pack $k$-point grid and a mesh cut-off of 1000 Ry for reciprocal / real-space
integrals were used for an accurate self-consistent 
convergence. We employed a double-$\zeta$ polarized 
basis set whose first-$\zeta$ PAOs had long radii of about 8 Bohr. All calculations were carried 
out using structural relaxations with a maximum force tolerance equal to 0.005 eV/\AA.
Interactions with the core electrons were included by a Troullier-Martins 
pseudopotential \cite{PhysRevB.43.1993}, obtained as described in Ref. \cite{Salva2015}. 
For the {\sc grogu} calculations,
a 100$\times$100$\times$1 Monkhorst-Pack $k$-grid was used and localized magnetic moments were 
projected onto the Cr $3d$-orbitals to avoid undesired contributions from non-magnetic 
orbitals \cite{PhysRevB.108.214418}.

\section{Results} \label{Sec_results}
First, we obtained the structural and electronic ground state of the three systems, CGS, CGT and CGST.
A summary of the results is given in Table \ref{Table_structure}. The lattice constant of CGST lies in between those of CGS and
CGT, as expected.
As of the magnetic structure, all these systems are ferromagnetic with a similar magnetic moment localized on each Cr atom, while all other atoms %in the considered structure 
are non-magnetic. The orbital contribution to the magnetic moments is below 1~\% an all cases.
%dominated by its spin contribution. MAE = $E_\parallel-E_\perp$
We define the magnetic anisotropy energy (MAE) as the difference of the self-consistently calculated total energy between in-plane and out-of-plane magnetizations, MAE = $E_\parallel-E_\perp$. Two-site $\frac{1}{2}\sum_{i\neq j}\left(J_{ij}^{xx}-J_{ij}^{zz}\right)$ and on-site $\sum_i\left(K_i^{xx}-K_i^{zz}\right)$ magnetic anisotropy contributions to the MAE can be obtained by choosing collinear in-plane (along the $x$-axis) and out-of-plane (along the $z$-axis) magnetizations in Hamiltonian \ref{HeisenbergEq2}. So the MAE is characterized by examining the $J_{cj}^{xx}-J_{cj}^{zz}$, $J_{cj}^{yy}-J_{cj}^{zz}$ and $K_c^{xx}-K_c^{zz}$ terms and the number of the nearest neighbors for each shell.
The MAE shows a preference of in-plane magnetization for CGS,
while CGT and CGST present an out-of-plane easy axis. This is consistent with experimental reports for CGT \cite{Gong2017}.  All compounds are semiconducting, with a band gap well below 1 eV.

\begin{table}[h]
\small
\caption{Summary of the structural, electronic and magnetic properties of CGS, CGT and CGST obtained from the {\sc SIESTA} calculations. Orbital and spin moments where computed by {\sc grogu}.} 
\begin{tabular*}{0.48\textwidth}{@{\extracolsep{\fill}}cccc}
 \hline
  & CGS & CGT & CGST \\
%Material & Lat. Const. (\AA) & $J^{H}_{c1}$& $J^{H}_{c2}$& $J^{H}_{c3}$&$J^{H}_{c4}$\\
 \hline
 Lattice constant (\AA) & 6.44 & 6.98 & 6.71 \\
 Spin moment ($\mu_B$) & 3.531 & 3.749 & 3.643 \\
 Orbital moment ($\mu_B$) & 0.008 & 0.022 & 0.0046 \\
 MAE (meV/Cr) & -0.05 & 0.80 & 0.15 \\
 Band gap (eV) & 0.8 & 0.2 & 0.3 \\
 \hline
\end{tabular*}
\label{Table_structure}
\end{table}
Next, we calculated the exchange parameters for these compounds, and in Table \ref{table_alldata}, we present the most relevant values up to the fifth nearest neighbors. The interactions are given in the local frame of reference shown in Fig. 
\ref{bondDM}(a), while the corresponding pairs of atoms $ij$ are labeled as depicted in Fig. \ref{Structure}(b). The negative
value of $J^{H}_{c1}$ indicates that the interaction between first neighbors is FM, which agrees well with experimental results 
for CGT \cite{Huang2017} and other theoretical results \cite{Chen2022,CGST}. The uniaxial on-site anisotropy energy, $K^{xx}_c-K^{zz}_c$, is
approximately $\sim$ 15 times larger in CGT than in CGS and 3 times larger than in CGST. This highlights the effect of SOC 
interaction in heavy atoms such as Te compared to the lighter atoms like Se. On-site anisotropy values for these materials show a 
preference of in-plane magnetization. However, the two-site exchange anisotropy $J_{cj}^{xx}-J_{cj}^{zz}$ and $J_{cj}^{yy}-J_{cj}^{zz}$,
also contributes to the total MAE, especially at first neighbors ($j=1$). This can be attributed to the exponential decay of the exchange interactions due to the semiconducting behavior of the materials. We find that it is this type of
magnetic anisotropy that leads to out-of-plane magnetization in CGT and CGST.
 
\begin{table}[h]
\small
\caption{Exchange parameters for CGS, CGT and CGST in the local reference frame given by Fig. \ref{bondDM}(a).  All quantities are given in meV units.
Atomic labels correspond to those defined in Fig. \ref{Structure} (b). Due to $C_3$ symmetry $K_c^{xx}=K_c^{yy}$. 
The values for $S_{c2}^{\alpha}$ are omitted in the table since their magnitudes are below 0.01 meV, with the exception of $S_{c2}^x$ in CGST 
where it is 0.04 meV. Isotropic exchange couplings beyond the fifth nearest neighbors are below $0.05$ meV. The number of first, second, third, fourth, and fifth nearest neighbors with respect to the $c$ atom are 3, 6, 3, 6 and 6, respectively.} %} 
\begin{tabular*}{0.48\textwidth}{@{\extracolsep{\fill}}cccc}
 \hline
 
 & CGS & CGT & CGST   \\
 \hline
     $K_c^{xx}-K_c^{zz}$ &-0.03&-0.47&-0.15\\
      \hline
     $J^{H}_{c1}$ &-0.86&-7.04&-4.38\\
     $J_{c1}^{xx}-J_{c1}^{zz}$&-0.01&-0.06&0.03\\
     $J_{c1}^{yy}-J_{c1}^{zz}$&0.09&0.72&0.58\\
     $S^{x}_{c1}$&-0.01&-0.02&-0.04\\
     $D_{c1}^y$&-&-&0.70\\
     $D_{c1}^z$&-&-&-0.19\\
 \hline
 $J^{H}_{c2}$ &0.28&0.19&0.72\\
 $J_{c2}^{xx}-J_{c2}^{zz}$&$<$0.01&-0.02&-0.04\\
 $J_{c2}^{yy}-J_{c2}^{zz}$&-0.01&-0.05&-0.04\\
 $D_{c2}^x$&-0.16&-0.19&-0.16\\
 $D_{c2}^y$&-&-&0.07\\
 $D_{c2}^z$&-0.19&-0.42&-0.25\\
 \hline
  $J^{H}_{c3}$ &0.30&-0.25&0.24\\
  $J_{c3}^{xx}-J_{c3}^{zz}$& $<$0.01&-0.11&-0.12\\
  $J_{c3}^{yy}-J_{c3}^{zz}$&-0.02&0.04&-0.05\\
  $S^{x}_{c3}$&-0.02&-0.14&-0.10\\
$D_{c3}^y$&-&-&-0.02\\
 $D_{c3}^z$&-&-&-0.28\\
 \hline
  $J^{H}_{c4}$ &0.21&0.19&0.24\\
 \hline
   $J^{H}_{c5}$ &0.34&0.37&red{0.39}\\
 \hline
\end{tabular*}
\label{table_alldata}
\end{table}

The only off-diagonal component of the symmetric exchange tensor that doesn't vanish by symmetry is
$S_{c1}^{x}$, but its value is very small as compared to the isotropic interactions, with no relevant differences among the three 
structures. Conversely, DM vectors at first nearest neighbors appear only for the Janus CGST compound due to the breaking of inversion symmetry. Of particular interest is the DM vector component $D_{c1}^{y}$ for CGST, with a value that is
approximately $\sim 16\%$ of the isotropic exchange $J^{H}_{c1}$ and agrees well with {\sc vasp} supercell 
calculations \cite{CGST}. %As shown in Fig. \ref{bondDM} (b), 
Remarkably, this component could induce the formation of
radial vortex magnetic patterns in CGST as shown in Fig. \ref{bondDM}(c).\\
\subsection{Biaxial Strain}
\begin{figure}[ht]
    \centering
    \includegraphics[width=\columnwidth]{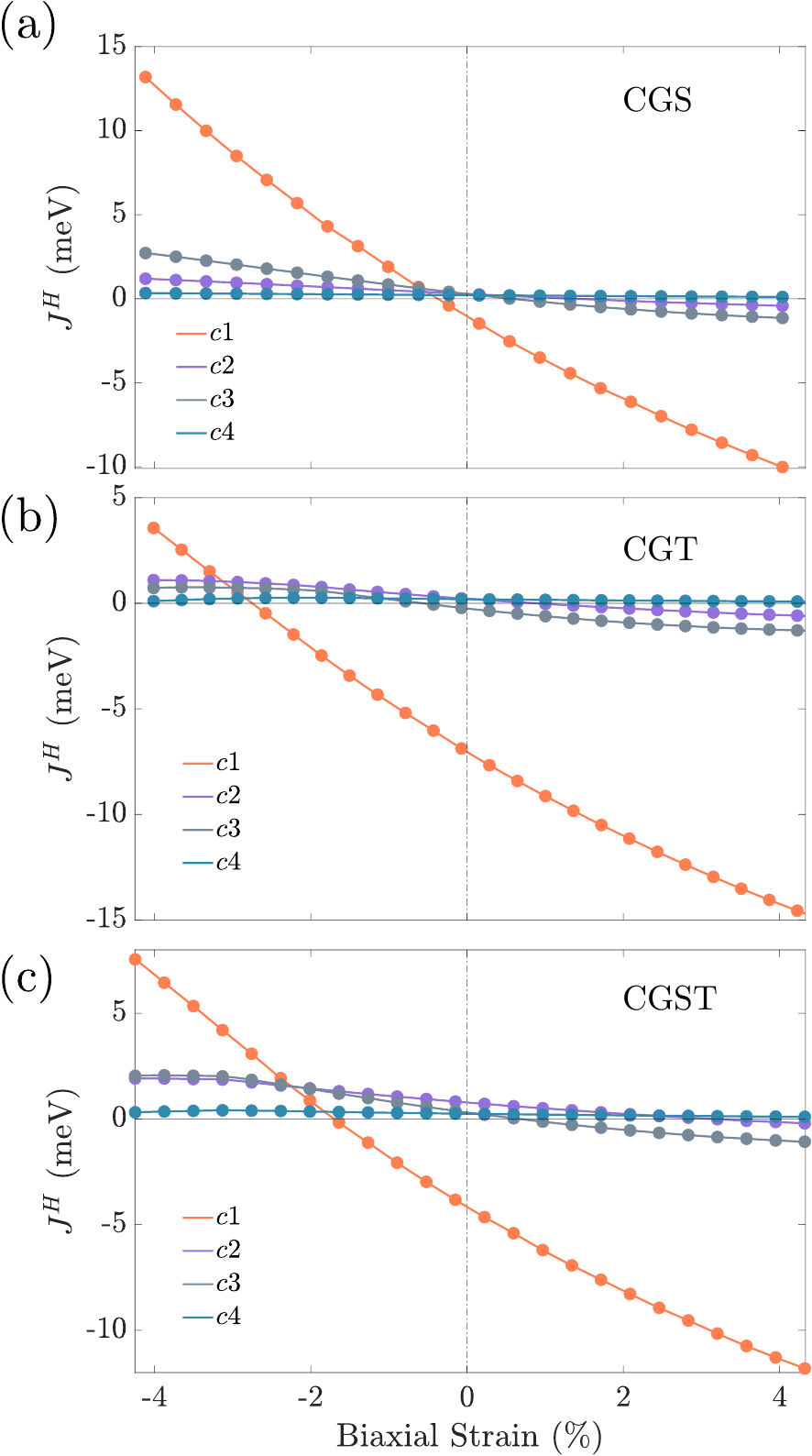}
    \caption{Isotropic exchange constants up to the fourth nearest neighbors as a function of the biaxial strain for (a) CGS, (b) CGT and (c) CGST. The labels $cj$ ($j$ = 1, 2, 3 and 4) refer to pairs of sites as depicted in Fig. \ref{Structure}(b).} %Vertical dashed black lines indicate the ground state structure.}
    \label{biaxial}
\end{figure}

%Once we have described 
After describing the magnetic interactions in the ground state, we focus on different effects
that could be applied experimentally to modify these interactions, eventually triggering a controllable change in the
magnetic response of those systems. We consider first the effect of homogeneous biaxial strain. This is simulated by relaxing
the structures for different fixed in-plane lattice constants. With respect to the ground state geometry, we look for strains in the range of $\pm 4 \%$. In this range,
the easy axis for the three materials remains unchanged, and the magnetic moments vary by approximately $2\%$, 
ensuring that the ground state multiplet is not affected by the strain.
\begin{comment}
Biaxial strain does not affect the symmetry of the system. As a consequence, we find that the most relevant
changes happen in the isotropic exchange constants. These are plotted in Fig. \ref{biaxial} as a function of strain.
The first neighbor interactions are the most sensitive on the strain because of the Goodenough-Kanamori-Anderson (GKA) 
mechanism \cite{PhysRev.100.564,KANAMORI195987,PhysRev.115.2}. This phenomenon is related to the angle
formed by consecutive Cr-X-Cr bonds (X = Se or Te), that increases with strain. When the angle passes the critical value of $\sim$ 90$^{\circ}$
an AFM/FM transition takes place. This mechanism explains that at short Cr-Cr distances the interaction 
at first nearest neighbors is AFM, but when the distance increases, a FM superexchange interaction mediated by Te or Se takes over.
The same mechanism is found in other magnetic systems such as 1T-CrTe$_2$ \cite{OteroFumega2020}, 
CrSiTe$_3$ \cite{CHEN201560} or other Janus magnets like Cr$_2$(X,Y)Te$_6$ (X, Y = Ge, Si, Sn) \cite{10.1063/5.0185859}.
\end{comment}
Biaxial strain does not affect the symmetry of the system. As a consequence, we find that the most relevant
changes happen in the isotropic exchange constants. These are plotted in Fig. \ref{biaxial} as a function of strain.
The first neighbor interactions are the most sensitive on the strain because of the Goodenough-Kanamori-Anderson (GKA) 
mechanism \cite{PhysRev.100.564,KANAMORI195987,PhysRev.115.2}. This phenomenon is related to the angle
formed by consecutive Cr-X-Cr bonds (X = Se or Te), that increases with strain. At short Cr-Cr distances, the interaction between Cr atoms is antiferromagnetic, driven by the Pauli exclusion principle and the presence of half-filled $t_{2g}$ orbitals in the Cr$^{3+}$ atoms. However, when the Cr-Cr distance increases and the angle bond is $\sim$ 90$^{\circ}$, a FM superexchange interaction mediated by the $p$ orbitals of Te or Se takes over \cite{Wang2022,Sun2018,Ren2019}. This implies that when the bond angle passes the critical value of $\sim$ 90$^{\circ}$, an AFM-to-FM transition occurs. This is due to the competition between the AFM direct exchange and the FM superexchange interactions, with the decay of the AFM direct exchange with Cr-Cr distance being more pronounced than that of the FM superexchange. The same mechanism is found in other magnetic systems such as 1T-CrTe$_2$ \cite{OteroFumega2020}, 
CrSiTe$_3$ \cite{CHEN201560} or other Janus magnets like Cr$_2$(X,Y)Te$_6$ (X, Y = Ge, Si, Sn) \cite{10.1063/5.0185859}.

The isotropic Heisenberg interactions for CGS are shown in Fig.~\ref{biaxial}(a) as a function of the lattice constant. For this compound, a compression of only 
$\sim$ 0.35$\%$ changes the sign of $J_{1c}^{H}$ to favor AFM interaction. Moreover, a larger compression of the order of 
$\sim$ 3$\%$ %on the lattice constant 
results in $J_{c1}^H\sim 8$ meV, which is about four times larger than the second strongest
isotropic coupling. This implies that CGS is a 2D magnet, where the ground state can be switched from FM to AFM experimentally through feasible biaxial compression.

As clear from Figs.~\ref{biaxial}(b) and (c), for the other two compounds larger compressions, $\sim2.7\%$ for CGT and $\sim1.6\%$ for CGST, are needed to 
have a sign change in $J_{c1}^{H}$. In case of a few strain values, in Table \ref{Table_angles1} we listed the angles between the Cr-Se/Te and Se/Te-Cr bonds connecting nearest neighbor Cr atoms. Regarding $J^H_{c1}$ for these strain values we find a good agreement with the GKA mechanism.
Note that for lattice constants below the FM-AFM transition for $J_{c1}^{H}$,   %After this transition, 
all isotropic 
constants up to the fourth nearest neighbors become positive (AFM coupling), indicating the loss of FM 
collinear magnetic order.

Beyond the effect on the isotropic exchange constants, we also noticed 
significant impacts of biaxial strain on the DM vectors, in particular, for ${\bf D}_{c2}$. In the considered range of strain, $D^x_{c2}$ varies between about -0.1 and -0.2 meV in CGT, while $D_{c2}^z$ varies between -0.1 and -0.2 meV in CGS and 
between -0.3 and -0.4 meV in CGT. %CGST

\begin{table}[h]
\small
\caption{Angle subtended  for the bonds 
Cr-X-Cr (X=Te and Se) in the three materials for some values of the biaxial strain. 
%In the third column it is shown 
For CGST it is shown the bond angles for Cr-Te-Cr / Cr-Se-Cr.} 
\begin{tabular*}{0.48\textwidth}{@{\extracolsep{\fill}}cccc}
 \hline
Biaxial Strain & CGS & CGT& CGST \\
 \hline
     -3$\%$ &  87.61$^{\circ}$ & 88.63$^{\circ}$ &  84.33$^{\circ}$ / 91.52$^{\circ}$ \\
     0$\%$ &  89.94$^{\circ}$  & 90.97$^{\circ}$ &  86.98$^{\circ}$/ 94.05$^{\circ}$\\
     +3$\%$ &  92.11$^{\circ}$  & 93.12$^{\circ}$ & 87.72$^{\circ}$ / 94.70$^{\circ}$  \\
 \hline
\end{tabular*}
\label{Table_angles1}
\end{table}

\subsection{Uniaxial Strain}

Next we investigate the effect of uniaxial strain on the exchange interactions.
We simulate a strain in the range of $\pm5\%$  applied only over the direction of $\vec{a}_1$, as defined
in Fig. \ref{Structure}(a). This strain breaks $C_{3}$ symmetry, therefore, the $c$1 and 
$c$1' bonds are no longer equivalent. This implies the lifting of degeneracy between the isotropic exchange parameters $J_{c1}^H$ and 
$J_{c1\text{'}}^H$, which can have different characteristics depending on the distance between Cr atoms and, consequently, on 
the Cr-X-Cr (X = Te, Se) bond angles. %To quantify the difference of the bond angles between the non-equivalent first nearest neighbors, we define $\Delta\alpha$ as 
The difference between the angles subtended by the $c$1 and $c$1' types of Cr-X-Cr bonds, %along the $c$1 and $c$1' bonds, 
$\Delta\alpha$,
is shown in Table. \ref{Table_angles2}. The values
of $\Delta\alpha$ are of the order of the differences between the angles shown in Table 
\ref{Table_angles1}, indicating that the effect of the GKA mechanism might be very different 
for the $c$1 and $c$1' bonds.

\begin{table}[ht!]
\small
\caption{Difference $\Delta\alpha$ between the angles subtended by the Cr-X-Cr bonds along the $c1$ and $c$1' nearest neighbors, when $\pm 3 \%$ uniaxial strain is applied over the $\vec{a}_1$ direction. For CGST, $\Delta\alpha$ is presented for the Cr-Te-Cr / Cr-Se-Cr bonds.} 
\begin{tabular*}{0.48\textwidth}{@{\extracolsep{\fill}}cccc}
 \hline
Uniaxial Strain & CGS & CGT& CGST \\
 \hline
     -3$\%$ & 2.30$^{\circ}$   & 2.42$^{\circ}$ & 2.18$^{\circ}$ / 2.48$^{\circ}$\\
     +3$\%$ & -2.33$^{\circ}$   &  -2.45$^{\circ}$& -2.22$^{\circ}$ / -2.31$^{\circ}$ \\
 \hline
\end{tabular*}
\label{Table_angles2}
\end{table}

\begin{figure}[h]
    \centering
    \includegraphics[width=0.9\columnwidth]{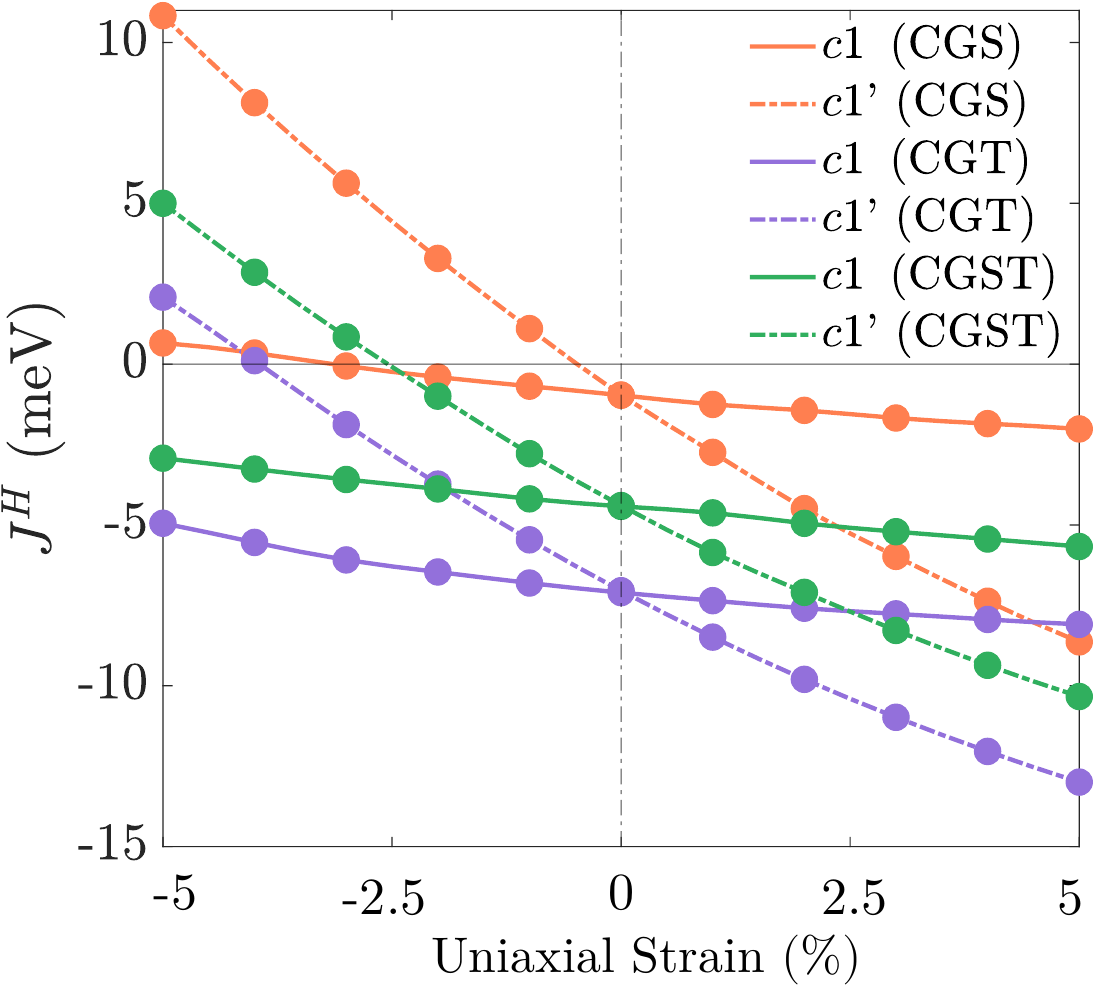}
    \caption{Nearest neighbor isotropic exchange couplings as a function of uniaxial strain over $\vec{a}_1$ for CGS (orange), CGT (purple) and CGTS (green). The interactions for the pairs split by the strain, labeled by $c$1 and $c$1', show remarkably different dependence on the strain. The labels $cj$ ($j$ = 1 and 1') refer to nearest neighbor pairs of sites as depicted in Fig. \ref{Structure}(b).}  
    \label{uniaxial}
\end{figure}

In Fig. \ref{uniaxial} the non-equivalent isotropic couplings for the nearest neighbors are shown as a 
function of the uniaxial strain. All the $J_{c1\text{'}}^{H}$ curves present similar slopes,
considerably larger than those of $J_{c1}^{H}$. This is a consequence of the $c$1' bond located
in the direction of the strain, and therefore being more sensitive to its magnitude.
Due to these different slopes, we find regions of compressive strain where $J_{c1\text{'}}^H$ changes sign
to AFM coupling but $J_{c1}^H$ remains FM. Most likely, these strain values would favor
a magnetic structure of FM coupled rows of Cr atoms in direction perpendicular to the strain, with AFM coupling between rows. The needed compression to obtain this transition is of only $\sim 0.5\%$ for CGS,
while a larger value of $\sim 2.5\%$ ($\sim 4\%$) is needed for CGST (CGT).

Focusing on the impact of uniaxial strain on the DM vectors we can note that most bonds retain their symmetries under uniaxial strain. The most obvious exception
is $c$2', since upon uniaxial strain the distance from site $1\text{'}$ to $c$ will differ from that between sites $1\text{'}$ to $2\text{'}$, 
and the $C_{2y}$ symmetry in the local frame of reference, see Fig. \ref{bondDM}(a) and Table~\ref{tablesymm}, gets broken. 
This does not happen for the $c$2 bond however. Therefore, $D_{c2\text{'}}^y$ is induced
by uniaxial strain. The magnitude of $D_{c2\text{'}}^y$ induced by $\pm5\%$ uniaxial strain is $\mp$0.1 meV 
($\mp$0.05 meV) that is $\sim 50\%$ ($16\%$) of $J^H_{c2\text{'}}$ for CGT (CGS). Recall that $C_{2y}$ symmetry 
was already broken in CGST and $D_{c2\text{'}}^y$ is different from zero even when uniaxial strain is not 
applied. No other relevant changes are observed for the relativistic exchange parameters, apart from those 
observed for biaxial strain.\\
From experimental point of view, compression through strain is difficult to achieve without avoiding buckling effects. However, compression values on the order of 0.2\% can be obtained \cite{Hui2013} and are close to the values needed to achieve the transition in CGS. Furthermore, to increase the compression, more sophisticated techniques may be used, such as  sulfur or oxygen substitution \cite{Li2018,Tang2020,Peto2018}, the introduction of vacancies \cite{Yang2019}, or other methods \cite{Peng2023}.

\subsection{Gate voltage}

\begin{figure}[htb]
    \centering
    \includegraphics[width=\columnwidth]{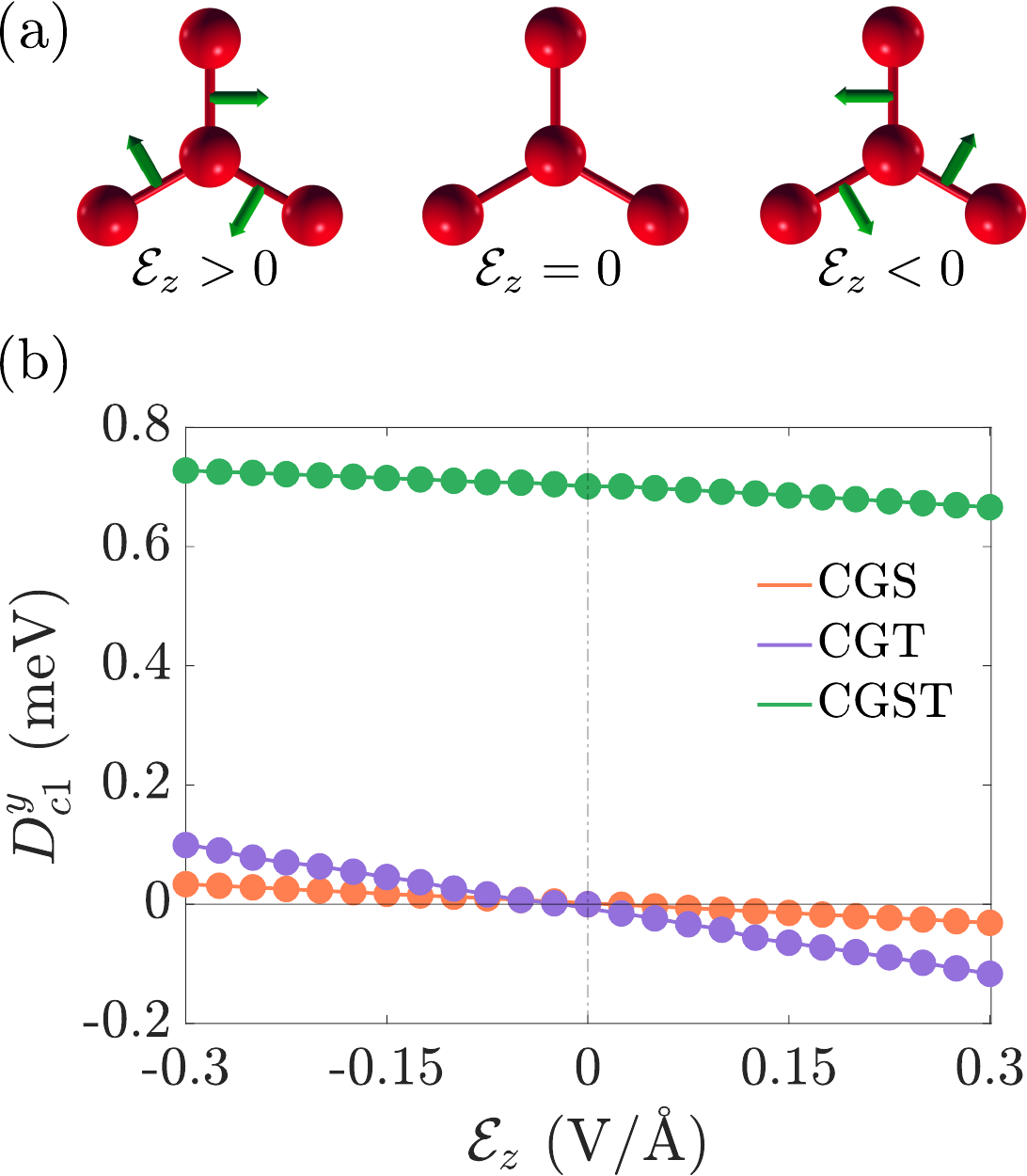}
    \caption{(a) Schematic drawing of the angular vortex pattern of the nearest neighbor DM  vectors depending on the sign of the applied electric field normal to a monolayer of CGS and CGT. (b)  $D_{c1}^y$ as a function of the electric field for CGS (orange), CGT (purple) and CGST (green). CGS and CGT curves cross the origin because of the inversion symmetry of the system when $\mathcal{E}_z=0$, unlike in the case of CGST. 
    }
    \label{Efield}
\end{figure}

We simulate the effect of gate voltage by including in our simulations an electric field $\mathcal{E}_z$ perpendicular to the monolayer. The range of electric field values we are using can be experimentally achieved with an STM tip \cite{Matvija2017}. 
The main effect of such an electric field is breaking the inversion symmetry 
of the CGS and CGT structures which induces DM interactions in the system \cite{PhysRevB.108.134430,PhysRevB.97.054416,Yang2018,doi:10.1126/sciadv.aav0265,Han2024}, while the isotropic and other exchange parameters are hardly affected by the gate voltage.
We find that $\mathcal{E}_z$ produces a spontaneous linear
induction in $D_{c1}^y$ for CGS and CGT which was zero in the absence of the field.
Inversion symmetry is already broken for CGST without electric field, however, a change in the magnitude of the DM vectors proportional to the electric field 
is also observed. Fig. \ref{Efield}(a) indicates that in CGS and CGT the chirality of the first nearest neighbor DM vectors, each parallel to the local $y$ axis, %by $D_{c1}^y$ 
is controlled by the sign of the electric field. In case of CGST, we 
find that $D_{c1}^y > 0$, so the chirality of the DM vectors is the one shown in the right picture in Fig. \ref{Efield}(a) in the investigated range of the electric field. 

Fig. \ref{Efield}(b) demonstrates the linear response in $D_{c1}^y$ to the applied electric field for
the three materials. A linear fit, performed by standard methods \cite{mat} to the expression $ D_{c1}^y 
(\mathcal{E}_z) = \mathcal{P}_{\text{X}}\cdot\mathcal{E}_z$  shows that 
$\mathcal{P}_{\text{CGS}}=-0.11$ e·\AA, $\mathcal{P}_{\text{CGT}}=-0.36$ e·\AA \ and $\mathcal{P}_{\text{CGST}}=-0.10$  
e·\AA. 
The induced $D_{c1}^y$ component can achieve 3$\%$ and 1.6$\%$  of $J^H_{c1}$ with $\mathcal{E}_z=0.3$ V/\AA\, for CGS and CGT, respectively.
For CGST, the linear response is small compared to the value in the absence of the field, 
and this induces a total magnitude change of $6\%$ of $D_{c1}^y (\mathcal{E}_z=0)$. 
Remarkably, in this compound $D_{c1}^y$ is between 15-16$\%$ of $J^H_{c1}$ for $\mathcal{E}_z\in[-0.3,0.3]$ V/\AA.

 The electrically induced DM interaction $D_{c1}^y$ is further enhanced by biaxial strain.
 We found that a strain of $3\%$ increases the magnitude of $D_{c1}^y$   
 by $\sim 60\%$ when $\mathcal{E}_z=0.3$ V/\AA\  is applied for CGS and CGT. Moreover, a biaxial compression of $3\%$ reduces the magnitude of $D_{c1}^y$ by $30\%$ and $46\%$ for CGS and CGT, respectively.

\begin{figure}[htb]
    \centering
    \includegraphics[width=0.9\columnwidth]{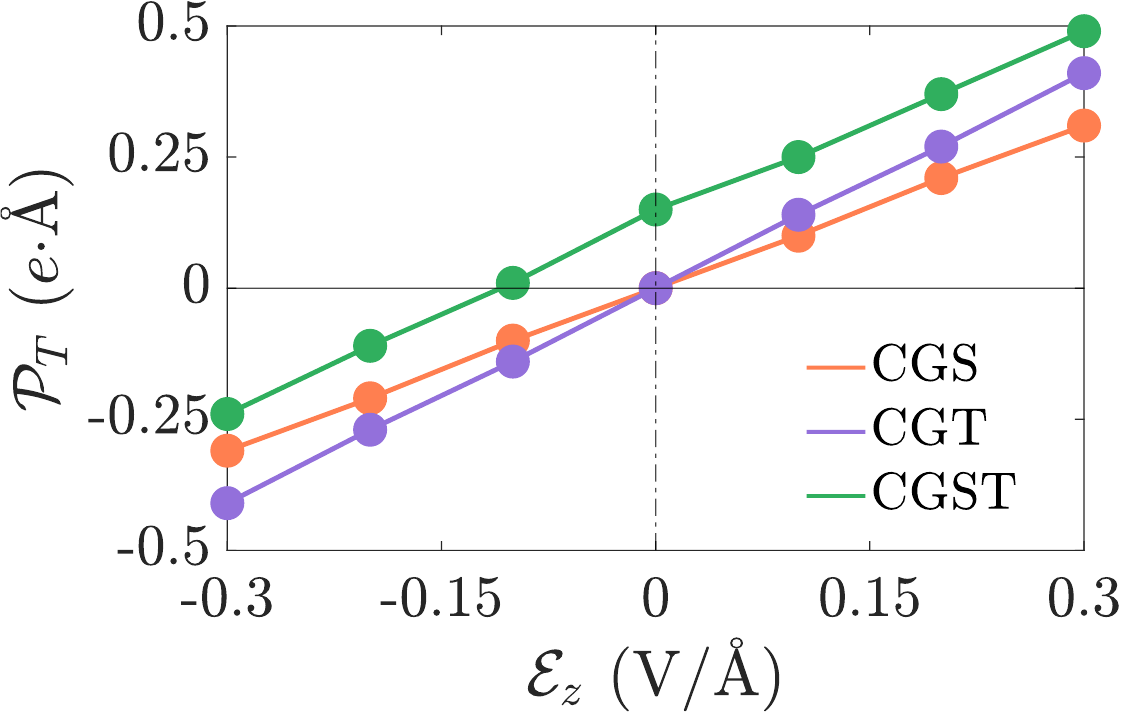}
    \caption{Total electric dipole moment $\mathcal{P}_T$ measured in $e$·\AA\ along the $z$-axis for the three different materials. In absence of electric field, CGST has a total dipole equal to 0.15 $e$·\AA\, that can be switched with electric fields below -0.1 V/\AA.
    }
    \label{Efield_pol}
\end{figure}

Lastly, we comment on the potential ferroelectric properties that can be found in Janus magnets \cite{Pang2023,PhysRevResearch.3.043011,Sun2022,Mahajan2023} like CGST. 
We find that the total electric dipole moment of CGS, CGT and CGST per unit cell follows the linear response relationship
\begin{equation}
   \mathcal{P}_T=\mathcal{P}_0 + \alpha \,{\cal E}_z 
\end{equation}
where $\alpha$ is the material's polarizability, see Fig. 
\ref{Efield_pol}. We find that the $z$-axis charge asymmetry in CGST produces a non-zero $\mathcal{P}_0=0.16$ e·\AA.\,  
The obtained $\mathcal{P}_0$ in this Janus magnet is about ten times larger than those obtained through bilayer stacking engineering \cite{PhysRevB.105.235445,bennett2024}, but several times smaller than those for perovskite-based compounds \cite{Ma2023}. The simultaneous presence of the electric dipole moment and magnetization in the same phase confirms the multiferroic behavior of CGST. However, since only a few Janus materials have been experimentally synthesized, no experimental measurements have yet been performed to confirm their multiferroic behavior. Furthermore,  a negative electric field larger in magnitude than  $\mathcal{E}_z=$0.1 V/\AA\ reverses the sign of $\mathcal{P}_T$, as shown in Fig. \ref{Efield_pol}. For CGS and CGT, $\mathcal{P}_0$ is zero. We find that the polarizability is similar for the three 
compounds. The proportionality of $D_{c1}^y$ 
with the electric field seen in Fig. \ref{Efield}(b) is consistent with the linear response $\mathcal{P}_T \propto {\cal{E}}_z$  
\cite{PhysRevB.97.054416,PhysRevMaterials.4.024405}.

\section{Summary and Conclusions} \label{Sec_conclusions}

We presented a computational study of the impact of symmetry breaking induced by different experimental
approaches on the magnetic properties of 2D vdW materials.
We calculated the tensorial exchange parameters for CrGeSe$_3$, CrGeT$_3$, and Cr$_2$Ge$_2$(Se,Te)$_3$ using the program package {\sc grogu} \cite{PhysRevB.108.214418} based on the relativistic version \cite{Udvardi2003} of the LKAG formalism \cite{liechtenstein87}. 
CGST exhibits ferroelectric properties and is a candidate to support magnetic textures due to DM interaction arising from the lack of inversion symmetry \cite{CGST}. Biaxial strain 
can trigger a FM/AFM transition among first neighbors with experimentally feasible compressions of approximately 0.4, 2, and 3$\%$ for CGS, CGST, and CGT, respectively. Uniaxial strain produces a difference between isotropic couplings for the $c1$ and $c1\text{'}$ types of nearest neighbor bonds, and reasonable values of 
compression ($\sim$ 0.5$\%$, 2.5$\%$ and 4$\%$ for CGS, CGST and CGT, respectively) results to %nearest neighbor 
isotropic couplings $J_{c1\text{'}}^H>0$ and $J_{c1}^H<0$. These couplings then promote a row-wise AFM texture of the Cr spins observed in other monolayer systems \cite{Spethmann2021,Elmers2023}.  
Uniaxial strain breaks $C_{2y}$ symmetry along the next nearest neighbor $c2$' bond and induces $D_{c2\text{'}}^y$ component that can achieve 50$\%$/16$\%$ of $J_{c2\text{'}}^H$ for CGT/CGS. A gate voltage perpendicular to the monolayer
breaks inversion symmetry for CGS and CGT, and $D_{c1}^y$ emerges in linear response to the electric field, forming a vortex pattern at the first nearest neighbors whose chirality is tuned by the sign of the electric field. Our computational results clearly demonstrate that relevant exchange parameters for the 2D vdW ferromagnets CGS, CGT, and CGST can be tuned by feasible external control to achieve magnetic phase transitions.

\section*{Acknowledgements}
G. M.-C., A. G.-F. and J. F. have been funded by Ministerio de Ciencia, Innovación y Universidades, Agencia 
Estatal de Investigación, Fondo Europeo de Desarrollo Regional via the grants PGC2018-094783 and PID2022-137078NB-I00,
and by Asturias FICYT under grant AYUD/2021/51185 with the support of FEDER funds. G. M.-C. has been supported by 
Programa ``Severo Ochoa'' de Ayudas para la investigación y docencia del Principado de Asturias.
This work was supported by the Ministry of Culture and Innovation and the National Research, Development and Innovation Office within the Quantum Information National Laboratory of Hungary (Grant No. 2022-2.1.1-NL-2022-00004)
and projects K131938, K142179. We thank the ”Frontline” Research Excellence Programme of the NRDIO, Grant No. KKP133827. This project has received funding from the HUN-REN Hungarian Research Network. { This project is supported by the TRILMAX Horizon Europe consortium (Grant No. 101159646).}

%\section*{Appendix: Moriya Rules} 
\section*{Appendix: Symmetry considerations}

Here we summarize %the Moriya rules \cite{PhysRev.120.91}, that is, 
the selection rules that occur
over the exchange constants of the Heisenberg Hamiltonian in Eq.~(\ref{Js+Ja}) due to the
invariance of the structure under certain symmetry operations. These rules are the extension of Moriya's rules for the DM vectors \cite{PhysRev.120.91}.
The symmetry operations are performed with origin at the center $\mathcal{C}$ of the bond between the atoms $A$ and $B$, as previously shown in Fig. \ref{bondDM}(a). The rules
are the following:

1. Inversion symmetry ($\Pi$) respect to $\mathcal{C}$:\\ $D^x_{AB}=D^y_{AB}=D^z_{AB}=0$.\\

2. $yz$-plane symmetry ($\sigma_{yz}$) passing through $\mathcal{C}$:\\ $S^y_{AB}=S^z_{AB}=D^x_{AB}=0$.\\

3. $zx$-plane symmetry ($\sigma_{zx}$) passing through $\mathcal{C}$:\\ $S^x_{AB}=S^z_{AB}=D^x_{AB}=D^z_{AB}=0$.\\

4. $xy$-plane symmetry ($\sigma_{xy}$) passing through $\mathcal{C}$:\\ $S^x_{AB}=S^y_{AB}=D^x_{AB}=D^y_{AB}=0$.\\

5. $C_2$ symmetry along $z$-axis ($C_{2z}$) passing through $\mathcal{C}$:\\ $S^x_{AB}=S^y_{AB}=D^z_{AB}=0$.\\

6. $C_2$ symmetry along $y$-axis ($C_{2y}$) passing through $\mathcal{C}$:\\ $S^x_{AB}=S^z_{AB}=D^y_{AB}=0$.\\

7. $C_2$ symmetry along $x$-axis ($C_{2x}$) passing through $\mathcal{C}$:\\ $S^y_{AB}=S^z_{AB}=D^y_{AB}=D^z_{AB}=0$.\\

8. $C_n$ ($n>2$) symmetry along $x$-axis ($C_{nx}$) passing through $\mathcal{C}$:\\ $S^x_{AB}=S^y_{AB}=S^z_{AB}=D^y_{AB}=D^z_{AB}=0$ and $J^{xx}_{AB}=J^{yy}_{AB}$.\\

The application of an electric field along the $z$-axis or Janus monolayers in the $xy$-plane break the following symmetries: $\Pi$, $\sigma_{xy}$, $C_{2x}$, and $C_{2y}$, therefore, the constraints following from rules 1., 4., 6. and 7. do not apply.

%\bibliography{biblio}% 

\end{document}